\documentclass[prd,showpacs,showkeys,nofootinbib,preprint]{revtex4-1}

\usepackage{amsmath,amsfonts,amssymb,wasysym}
\usepackage{bm}
\usepackage{float} 
\usepackage{tikz}
\usepackage{pgfplots}
\usepackage{hyperref}

\def\lambdabar{\protect\@lambdabar}
\def\@lambdabar{%
\relax
\bgroup
\def\@tempa{\hbox{\raise.73\ht0
\hbox to0pt{\kern.25\wd0\vrule width.5\wd0
height.1pt depth.1pt\hss}\box0}}%
\mathchoice{\setbox0\hbox{$\displaystyle\lambda$}\@tempa}%
{\setbox0\hbox{$\textstyle\lambda$}\@tempa}%
{\setbox0\hbox{$\scriptstyle\lambda$}\@tempa}%
{\setbox0\hbox{$\scriptscriptstyle\lambda$}\@tempa}%
\egroup}

\begin{document}

\title{Existence of a system of discrete volume-localized quantum levels for charged fullerenes}

\author{Rafael V. Arutyunyan}
\email{arut@ibrae.ac.ru}
\affiliation{Nuclear Safety Institute, Russian Academy of Sciences, B. Tulskaya 52, 115191 Moscow, Russia}

\author{Petr N. Vabishchevich}
\email{vab@ibrae.ac.ru}
\affiliation{Nuclear Safety Institute, Russian Academy of Sciences, B. Tulskaya 52, 115191 Moscow, Russia}

\author{Yuri N. Obukhov}
\email{obukhov@ibrae.ac.ru}
\affiliation{Nuclear Safety Institute, Russian Academy of Sciences, B. Tulskaya 52, 115191 Moscow, Russia}

\begin{abstract}
In the framework of a simple physical model, we demonstrate the existence of a system of discrete short-lifetime quantum levels for electrons in the potential well of the self-consistent field of charged fullerenes and onion-like structures. For electrons, in the case of positively charged fullerenes and onion-like structures, combining analytic and numeric considerations we find that the energy of the volume-localized levels ranges from 1 eV to 100 eV. 
\end{abstract}
\maketitle

\section{Introduction}

Fullerenes represent one of allotropes of carbon, along with graphite, diamond, amorphous carbon, nanotubes and graphene. Following the earlier theoretical predictions, the first fullerene $C_{60}$ molecule was experimentally discovered in the 1980-ies \cite{Kroto:1985,Kraetschmer:1990} as a nanometer-size hollow spherical structure of 60 carbon atoms located at the vertices of a truncated icosahedron. Subsequently, the production of fullerenes in large quantities was developed and the fullerene nanotubes and many other fullerenes were discovered, such as $C_{20}$, $C_{70}$ and even lager structures. This gave a start to an explosive growth of research in the area of nanoscience, the historic development and the current status of which can be found in the numerous reviews \cite{Eletski:1995,Hirsch:2005,Sattler:2010,Gogotsi:2010,Campbell}. 

During the recent time, the properties of charged fullerenes have been actively experimentally and theoretically investigated \cite{Radi:1990,Troullier:1992,Yabana:1993,Brenac:1999,Bhardwaj:2003,Jensen:2004,Zettergren:2002,Zettergren:2004,Diaz:2005a,Diaz:2005b,Sahnoun:2006a,Sahnoun:2006b,Iroshnikov:2006,Iroshnikov:2007,Kono:2007,Madjet:2008,Nascimento:2011,Polozkov:2005,Polozkov:2013,Luders:2009,Liu:2018}. A considerable number of works are devoted to the study of their stability (lifetime), mechanisms for their charging and decay \cite{Wang:2011}.

The present paper is devoted to the discussion of the structure of the electronic spectrum of the charged fullerenes. Simple models are used to show the existence of the volume-localized discrete quantum levels for the usual fullerene and for the onion-like structures \cite{preprint}. Here we confine our attention to the case of the positively charged fullerenes. 

Basic notations are as follows: $m_e$ and $e$ are electron's mass and the absolute value of electron charge, $\lambdabar_e = {\frac {\hbar}{m_ec}}$ is the electron Compton length, $\varepsilon_0$ is the electric constant of vacuum; $a_0 = {\frac {4\pi\varepsilon_0\hbar^2} {m_e e^2}}$ is the Bohr radius, $\alpha = {\frac {e^2}{4\pi\varepsilon_0\hbar c}}$ is the fine structure constant.

\section{Discrete volume-localized levels: qualitative preliminaries}

As a preliminary step, let us formulate the corresponding quantum-mechanical spectral problem. With an account of the spherical symmetry of a fullerene, for the wave function we use the standard ansatz $\psi(r,\vartheta,\varphi) = R(r)Y_{lm}(\vartheta,\varphi)$, with the spherical harmonics $Y_{lm}$, and recast the spherically symmetric Schr\"odinger equation \cite{Landau} into a second order differential equation 
\begin{equation}\label{4}
 \frac{d^2\chi}{d r^2} - \frac{l(l+1)}{r^2}\,\chi + {\frac {2m_e}{\hbar^2}}(E - U(r)) \chi = 0
\end{equation} 
for the function $\chi(r) = r R(r)$, $0 \leq r < \infty$, under the boundary conditions
\begin{equation}\label{5}
 \chi(0) = 0, \quad \chi(\infty) = 0.
\end{equation} 
The form of solution is determined by the potential $U(r)$.

We will discuss the energy levels of an electron by starting from a simple model potential, and then move on to more complicate form of $U$. In the simplest model, one can describe a charged fullerene by the potential of a sphere with a constant surface charge density:
\begin{equation}\label{1}
  U(r) = -\,Z\,\Phi(r),\quad \Phi = {\frac {e}{4\pi\varepsilon_0}}
  \times\left \{ \begin{array}{cc}
  {\displaystyle {\frac 1R} },  &  r \leq R , \\
   & \\
  {\displaystyle {\frac 1r} },  &  r > R , \\
\end{array}\right . 
\end{equation} 
where $Z =  Ne$ is a positive charge, and $R = R_f$ is the fullerene radius. Our attention will be mainly confined to the $C_{60}$ fullerene, when $R_f = 6.627 a_0$. 

A first very approximate estimate of the energy levels can be obtained by replacing (\ref{1}) with a spherical rectangular well of the depth $U_0 = {\frac {Ze}{4\pi\varepsilon_0R_f}}$. Inside such a well ($0\leq r \leq R_f$), a non-normalized solution of the Schr\"odinger equation (\ref{4}) is described by the spherical Bessel function $\chi = j_l\left(\xi r/R_f\right)$ which satisfies the boundary condition at zero $\chi(0) = 0$, whereas outside the well ($R_f < r < \infty$) a solution that satisfies $\chi(\infty) = 0$ is given by the spherical Hankel function $\chi = h_l\left(i\eta r/R_f\right)$. The parameters $\eta$ and $\xi$ are algebraically related,
\begin{equation}
\xi^2 + \eta^2 = {\frac{2m_eU_0R_f^2}{\hbar^2}} = 2\alpha {\frac Ze} {\frac {R_f}{\lambdabar_e}},\label{U}
\end{equation}
and they determine discrete energy levels via  
\begin{equation}\label{2}
 E = - \,{\frac {\hbar^2 \eta^2}{2m_e R_f^2}} = -\,U_0 + {\frac {\hbar^2 \xi^2}{2m_e R_f^2}}.
\end{equation}
The values of parameters $\eta$ and $\xi$ are fixed by the continuity condition of the wave function at $r = R_f$. For $l = 0$, this yields 
\begin{equation}\label{3}
 \eta = - \,\xi \cot \xi\,.
\end{equation}
It is worthwhile to notice that the right-hand side of (\ref{U}) is essentially greater than 1, and already for a single charged $C_{60}^{+1}$ we have ${\frac{2m_eU_0R_f^2}{\hbar^2}} = 13.25$. As a result, for highly charged fullerenes ($Z \apprge 10e$) one can use an approximation of a very deep well $U_0 \gg {\frac {\hbar^2}{2m_eR_f^2}}$, deriving the energy levels from a condition of the vanishing of the wave function at the boundary: $j_l\left(\xi\right) = 0$. 

The characteristic feature of the corresponding wave functions is that they obviously describe the volume-localized states which are basically confined to the central part of the potential, i.e., to the inner region of the fullerene $r \leq R_f$. The number of such states increases for the potential well becoming deeper, which happens when the charge $Z$ of the fullerene grows. 

An interesting issue is actually how high is the value of an electric charge that a fullerene can carry? In practice, one can multiply ionize $C_{60}$ with the help of the highly charged ions, fast electrons, or photons \cite{Campbell}. Experimentally, charged fullerenes in the range of $Z = 0, \cdots , 9e$, \cite{Jensen:2004}, and even up to $Z = 10e$, \cite{Brenac:1999}, were produced in collisions of a beam of $C_{60}$ with a beam of highly ionized Xe atoms; such charged fullerenes are stable on a time scale of several $\mu$s. The highest value $Z = 12e$ was observed for a charged fullerene (with the lifetime of of order of a $\mu$s) ionized by intense short infrared laser pulses \cite{Bhardwaj:2003}. The theoretic analysis of the Coulomb stability of highly charged fullerenes \cite{Zettergren:2002,Zettergren:2004} predicted the limiting value $Z = 18e$ on the basis of a conducting sphere model, whereas the existence of $Z = 14e$ was established theoretically \cite{Diaz:2005a,Diaz:2005b,Sahnoun:2006a,Sahnoun:2006b} by means of the density functional theory. However, the predicted lifetime falls drastically --by ten orders-- when $Z$ increases from 11 to 14.

\section{Analytic potential for a charged fullerene}

\begin{table}
\caption{Electron energy levels for model potential (\ref{6}). [Notation: $n$ -- level number, $l$ -- angular quantum number, $i$ -- radial quantum number].}\label{t-11}
  \begin{tabular}{ccccc|ccccc}
\hline
    $n$  & $l$ & $i$ & $E$ (au) & & & $n$  & $l$ & $i$ & $E$ (au) \\
  \hline
  1    &    0    &    1   &     $-0.93858$  & & &
  7    &    6    &    1   &     $-0.45503$   \\  
  2    &    1    &    1   &     $-0.91502$  & & &
  8    &    7    &    1   &     $-0.29839$   \\    
  3    &    2    &    1   &     $-0.86809$  & & &
  9    &    8    &    1   &     $-0.12197$   \\    
  4    &    3    &    1   &     $-0.79812$  & & &
 10    &    0    &    2   &     $-0.05641$   \\    
  5    &    4    &    1   &     $-0.70558$  & & &
 11    &    1    &    2   &     $-0.02985$   \\   
  6    &    5    &    1   &     $-0.59101$  & & &
 12    &    2    &    2   &     $-0.00130$   \\
\hline
  \end{tabular}
\end{table} 

The model above provides a rather simplified description in the sense that it does not take into account the actual physical structure of a fullerene. A more realistic potential $U(r)$ can be constructed in the framework of the jellium model \cite{Rubio:1994,Ivanov:2001,Ivanov:2003,Belyaev:2009,Verkhovtsev:2012,Baltenkov:2015} as a sum of the positive contribution of the carbon atom's nuclei located on the spherical surface of the fullerene radius $R_f$ and the negative contribution of the electron clouds. The resulting potential is attractive and it has a cusp-shape form with the clear localization in the thin spherical shell. For $C_{60}$, the corresponding Lorentz-bubble potential reads
\begin{equation}\label{6}
 U(r) = - \,\frac{{\frac {\hbar^2}{m_e}}V}{(r-R)^2 + d^2}\,,
\end{equation} 
where the parameter $V$ determines the depth, $d$ the width, and $R$ the position. In the self-consistent spherical jellium model based on the Kohn-Sham equations, these parameters are fixed \cite{Baltenkov:2015} to the values
\begin{equation}\label{VRd}
 V = 0.711,  \quad R = 6.627\,a_0,  \quad d = 0.610\,a_0.
\end{equation}

\begin{table}
\caption{\label{t-12}Electron energy levels for charged fullerene potential (\ref{7}) with $Z = 1e$. [Notation: $n$ -- level number, $l$ -- angular quantum number, $i$ -- radial quantum number].}
  \begin{tabular}{ccccc|ccccc}
  \hline
   $n$  & $l$ & $i$ & $E$ (au)  & & & $n$  & $l$ & $i$ & $E$ (au) \\
  \hline
    1    &     0    &    1   &    $-1.08421$  & & &
    7    &     6    &    1   &    $-0.59927$  \\
    2    &     1    &    1   &    $-1.06058$  & & &
    8    &     7    &    1   &    $-0.44209$  \\
    3    &     2    &    1   &    $-1.01352$  & & &
    9    &     8    &    1   &    $-0.26499$  \\
    4    &     3    &    1   &    $-0.94336$  & & &
   10    &     0    &    2   &    $-0.18757$  \\
    5    &     4    &    1   &    $-0.85057$  & & &
   11    &     1    &    2   &    $-0.14965$  \\
    6    &     5    &    1   &    $-0.73566$  & & &
   12    &     2    &    2   &    $-0.10580$  \\  
  \hline
  \end{tabular} 
\end{table} 

\begin{figure}
  \begin{center}
    \includegraphics[width=1.\linewidth] {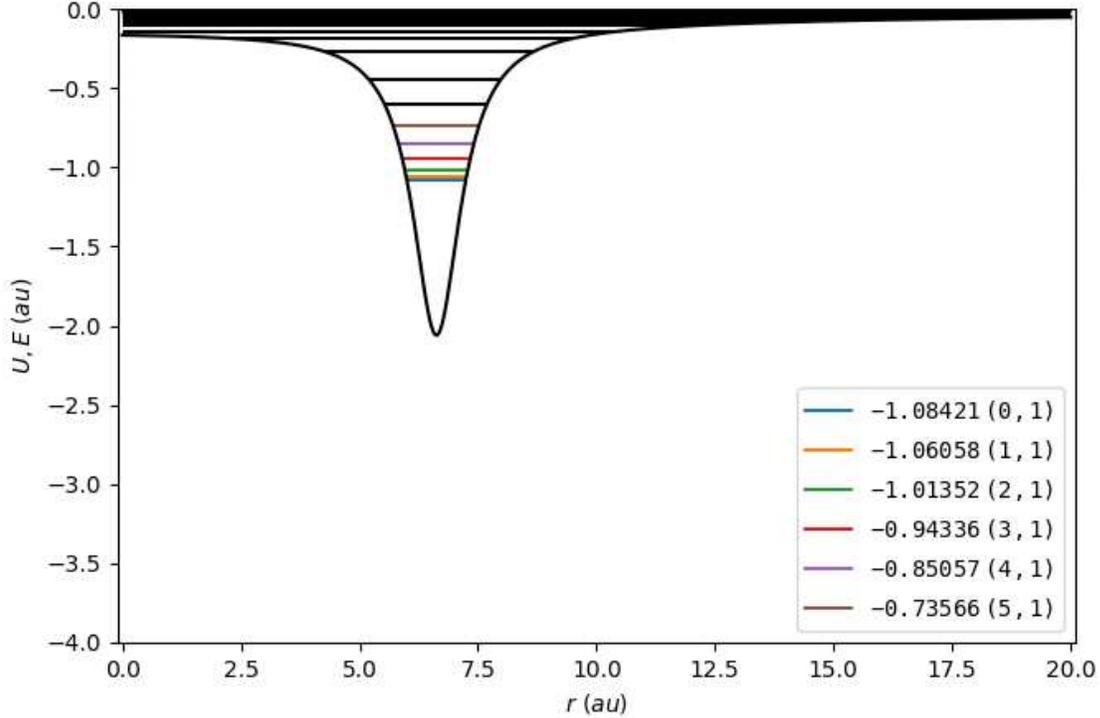}
	\caption{Spectrum for  charged fullerene potential (\ref{7}) with $Z = 1e$.}
	\label{f-17}
  \end{center}
\end{figure} 

In order to test the computational methods used in this paper, we found the energy levels $E$ for different values of the angular quantum number $l$ by integrating the Schr\"odinger equation (\ref{4}) numerically. The results, see Table~\ref{t-11}, reproduce the findings of \cite{Baltenkov:2015}. In contrast to the volume-localized feature of the wave functions for the model (\ref{1}), the states for the potential (\ref{6}) mostly have a typical surface-localized behavior. At the center of a fullerene, the value $U(0) = -\,0.016\,$au $= -\,0.44\,$eV is only slightly below zero, and hence only few discrete levels with the negative energy higher than that value correspond to the volume-localized states. 

Coming to the case of a charged fullerene, let us now modify the Lorentz-bubble potential (\ref{6}) by including the contribution of the charged spherical surface (\ref{1}). The generalization of the potential (\ref{6}) for a charged fullerene model then reads
\begin{equation}\label{7}
 U(r) = -\,\frac{{\frac {\hbar^2}{m_e}}V}{(r-R)^2 + d^2} - Z\Phi(r),
\end{equation} 
where $Z$ is the charge of the fullerene.

With such a modification, the central part of the potential deepens, so that $U(0) = -\,0.17\,$au $= -\,4.5\,$eV already for $Z = 1e$, whereas $U(0) = -\,0.77\,$au $= -\,20.9\,$eV for $Z = 5e$, and $U(0) = -\,1.52\,$au $= -\,41.5\,$eV for $Z = 10e$. As a result, there are two types of wave functions for the modified potential (\ref{7}): the lower-energy states are distinctly surface-localized, whereas the higher energy levels correspond to the volume-localized quantum states. 

\begin{figure}
  \begin{center}
    \includegraphics[width=1.\linewidth] {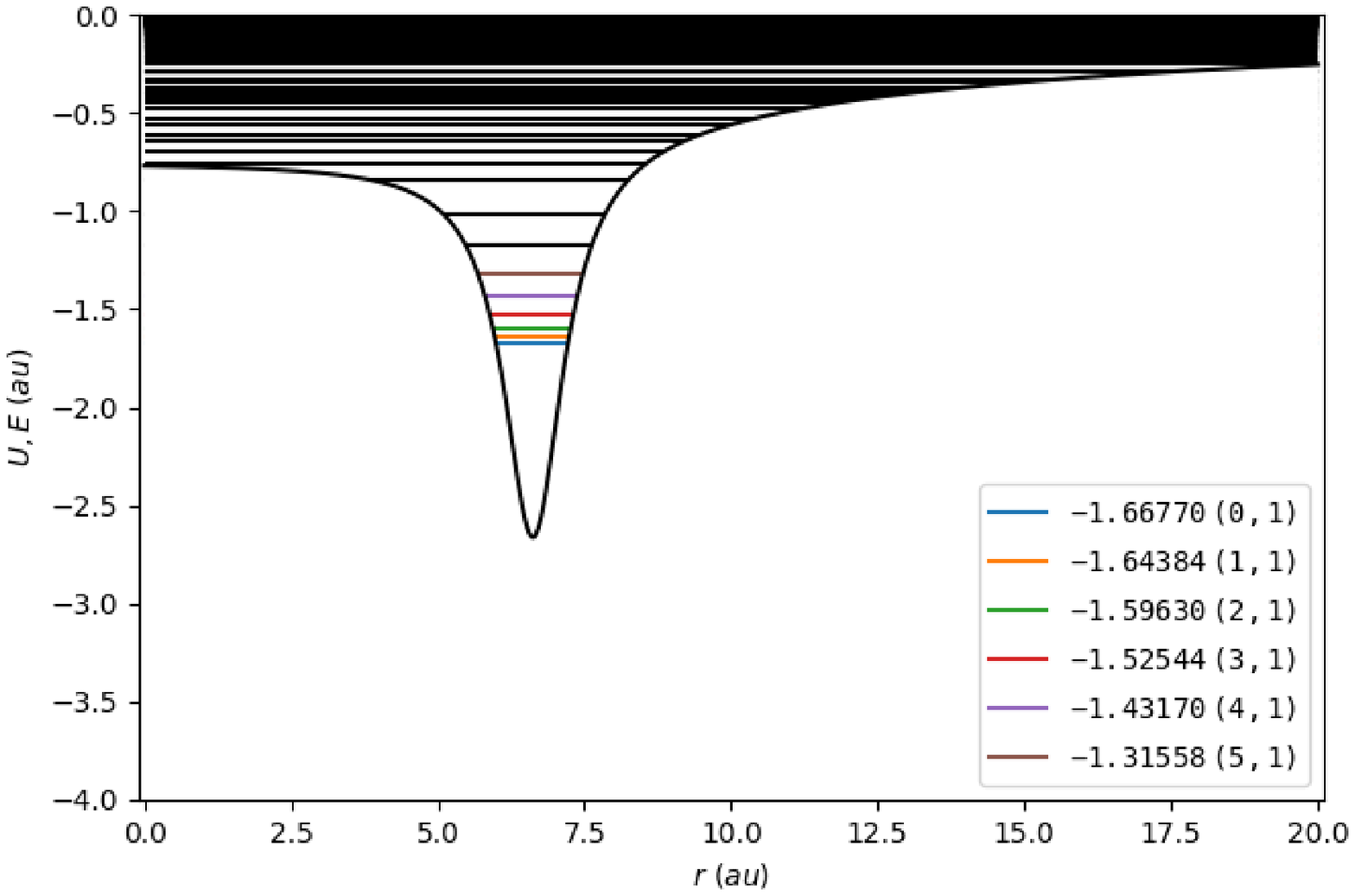}
	\caption{Spectrum for  charged fullerene potential (\ref{7}) with $Z = 5e$.}
	\label{f-19}
  \end{center}
\end{figure}

\begin{figure}
  \begin{center}
    \includegraphics[width=1.\linewidth] {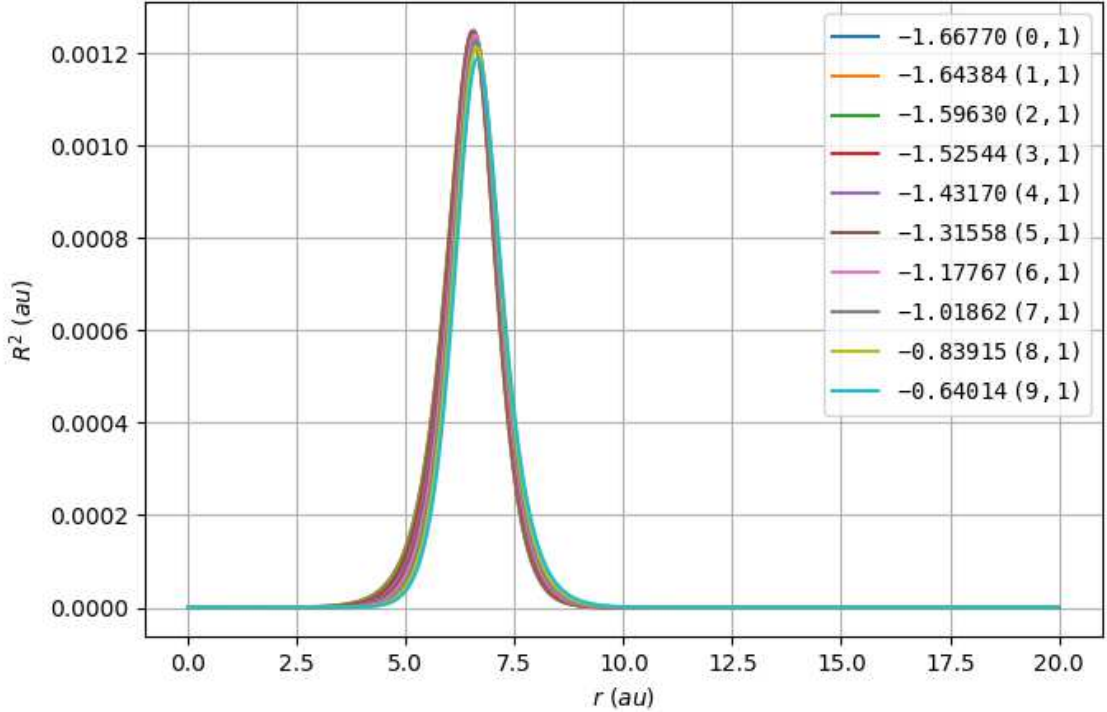}
	\caption{Surface-localized wave functions for charged fullerene potential (\ref{7}) with $Z = 5e$.}
	\label{surf}
  \end{center}
\end{figure}

\begin{figure}
  \begin{center}
    \includegraphics[width=1.\linewidth] {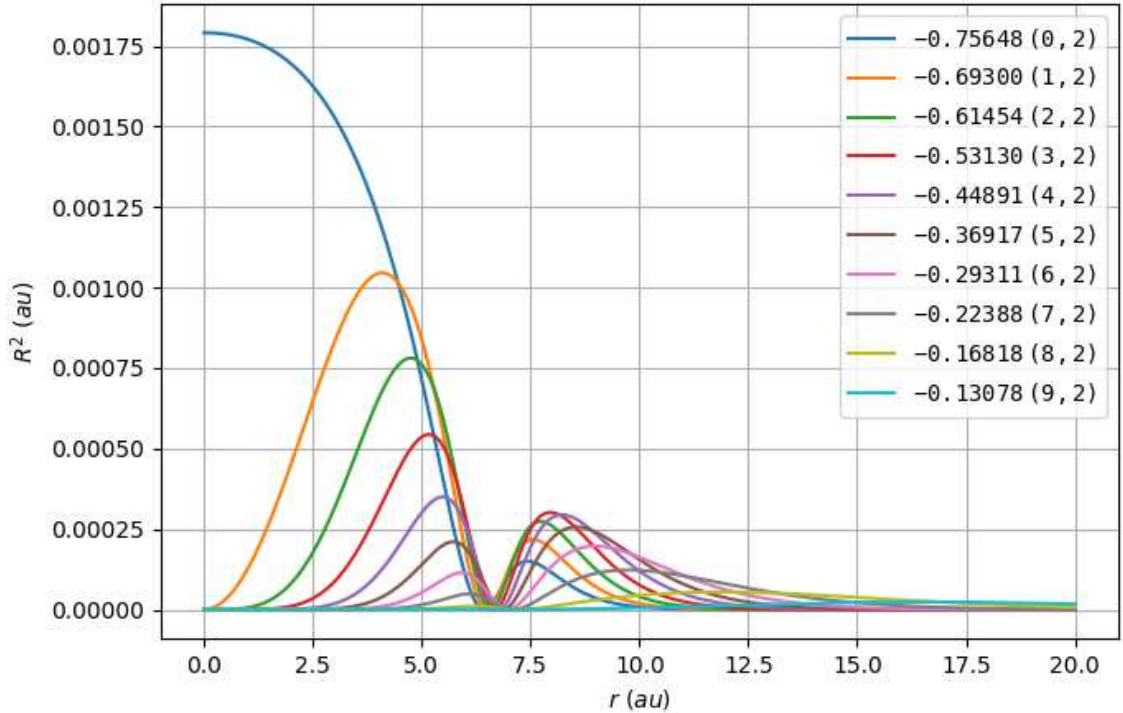}
	\caption{Volume-localized wave functions for charged fullerene potential (\ref{7}) with $Z = 5e$.}
	\label{vol}
  \end{center}
\end{figure}

\begin{figure}
  \begin{center}
    \includegraphics[width=1.\linewidth] {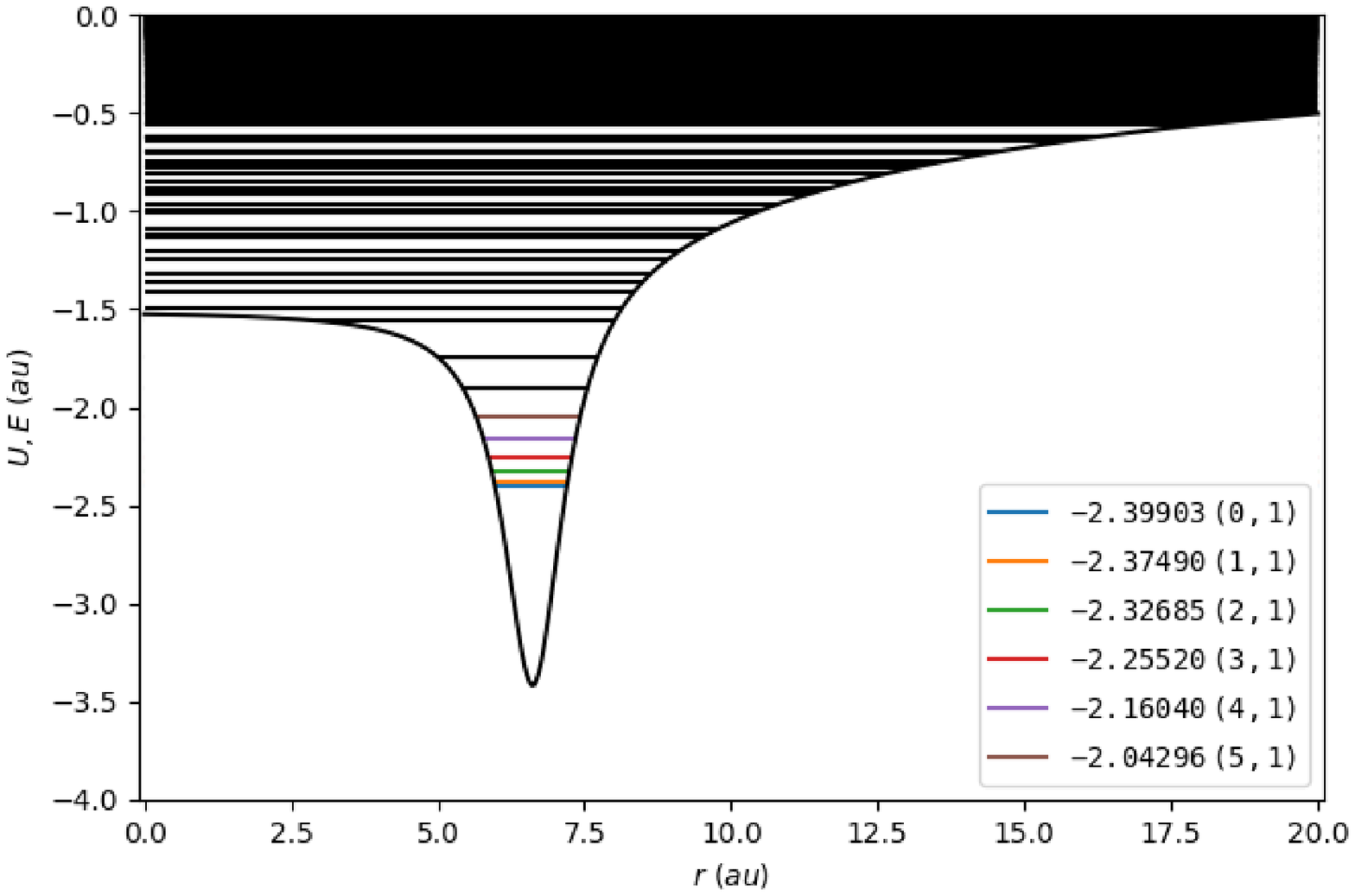}
	\caption{Spectrum for  charged fullerene potential (\ref{7}) with $Z = 10e$.}
	\label{f-20}
  \end{center}
\end{figure}

We find the discrete quantum energy levels of an electron in the potential (\ref{7}) by integrating the Schr\"odinger equation (\ref{4}) numerically. The corresponding results for different values of the fullerene charge $Z$ are presented in Tables~\ref{t-12}-\ref{t-15} and Figs.~\ref{f-17}-\ref{f-20}. We limit ourselves to the first 12 eigenvalues. 

\begin{table}
\caption{Energy levels for  charged fullerene potential (\ref{7}) with $Z = 5e$. [Notation: $n$ -- level number, $l$ -- angular quantum number, $i$ -- radial quantum number].}\label{t-14}
  \begin{tabular}{ccccc|ccccc}
  \hline
   $n$  & $l$ & $i$ & $E$ (au)  & & & $n$  & $l$ & $i$ & $E$ (au) \\
  \hline
         1    &    0    &    1   &    $-1.66770$  & & &
         7    &    6    &    1   &    $-1.17767$  \\
         2    &    1    &    1   &    $-1.64384$  & & &
         8    &    7    &    1   &    $-1.01862$  \\
         3    &    2    &    1   &    $-1.59630$  & & &
         9    &    8    &    1   &    $-0.83915$  \\
         4    &    3    &    1   &    $-1.52544$  & & &
        10    &    0    &    2   &    $-0.75648$  \\
         5    &    4    &    1   &    $-1.43170$  & & &
        11    &    1    &    2   &    $-0.69300$  \\
         6    &    5    &    1   &    $-1.31558$  & & &
        12    &    9    &    1   &    $-0.64014$  \\  
  \hline
  \end{tabular} 
\end{table} 

\begin{table}
\caption{Electron energy levels for  charged fullerene potential (\ref{7}) with $Z = 10e$. [Notation: $n$ -- level number, $l$ -- angular quantum number, $i$ -- radial quantum number].}\label{t-15}
  \begin{tabular}{ccccc|ccccc}
  \hline
   $n$  & $l$ & $i$ & $E$ (au)  & & & $n$  & $l$ & $i$ & $E$ (au) \\
  \hline
   1    &    0   &     1   &    $-2.39903$  & & &
   7    &    6   &     1   &    $-1.90343$  \\
   2    &    1   &     1   &    $-2.37490$  & & &
   8    &    7   &     1   &    $-1.74240$  \\
   3    &    2   &     1   &    $-2.32685$  & & &
   9    &    8   &     1   &    $-1.56053$  \\
   4    &    3   &     1   &    $-2.25520$  & & &
  10    &    0   &     2   &    $-1.49334$  \\
   5    &    4   &     1   &    $-2.16040$  & & &
  11    &    1   &     2   &    $-1.41574$  \\
   6    &    5   &     1   &    $-2.04296$  & & &
  12    &    9   &     1   &    $-1.35854$  \\  
  \hline
  \end{tabular} 
\end{table} 

It is known that for the choice of parameters (\ref{VRd}), the depth of the potential well (\ref{6}) is too large and the calculated energy levels are not in agreement with the experimental value of the electron affinity 2.65 eV for $C_{60}$. This issue was discussed in \cite{Madjet:2008,Belyaev:2009,Verkhovtsev:2012,Baltenkov:2015} for different model potentials in order to bring theory to a better agreement with experiment. By making use of the improved parameter set $V = 0.1104, R = 6.665\,a_0, d = 0.5\,a_0$, one finds an essentially shallower potential with the value at the center $U(0) = -\,0.0025\,$au that correctly reproduces the 2.65 eV detachment energy for the neutral $C_{60}$ molecule \cite{Baltenkov:2015}. For such an improved parameter choice, the central part of the potential (\ref{7}) for charged fullerene is shifted to $U(0) = -\,1.5\,$au $= -\,4.1\,$eV for $Z = 1e$, to $U(0) = -\,0.75\,$au $= -\,20.5\,$eV for $Z = 5e$, and to $U(0) = -\,1.5\,$au $= -\,40.8\,$eV for $Z = 10e$. Notice that with an increase of the electric charge $Z$, the form of the model potential changes very little when the parameters (\ref{VRd}) are replaced by an improved parameter set. As a result, the numeric analysis of the electronic spectrum for charged fullerenes with the improved set of parameters of the model potential confirms the existence of the volume-localized states, predicting slightly shifted theoretic values of the corresponding energy levels. 

\begin{figure}
  \begin{center}
    \includegraphics[width=1.\linewidth] {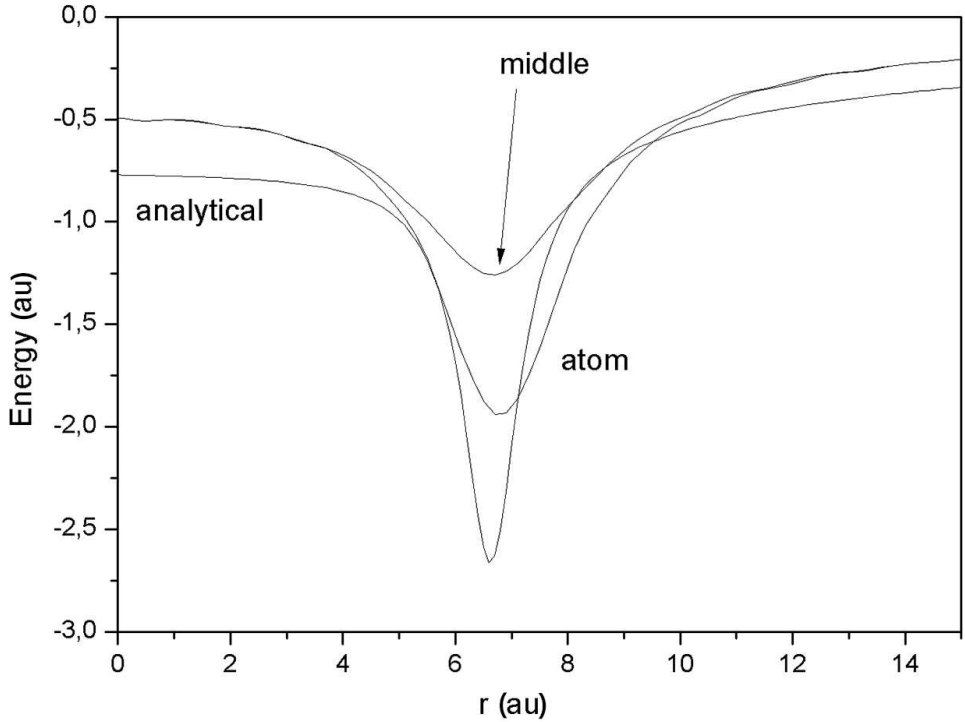}
	\caption{Comparison of the analytical potential of the charged fullerene (\ref{7}) for $Z = 5e$ with the DFT result: along the radial direction through carbon's atom (``atom'' curve) and through the center between neighboring atoms (``middle'' curve).}
	\label{dft}
  \end{center}
\end{figure}

\begin{figure}
  \begin{center}
    \includegraphics[width=1.\linewidth] {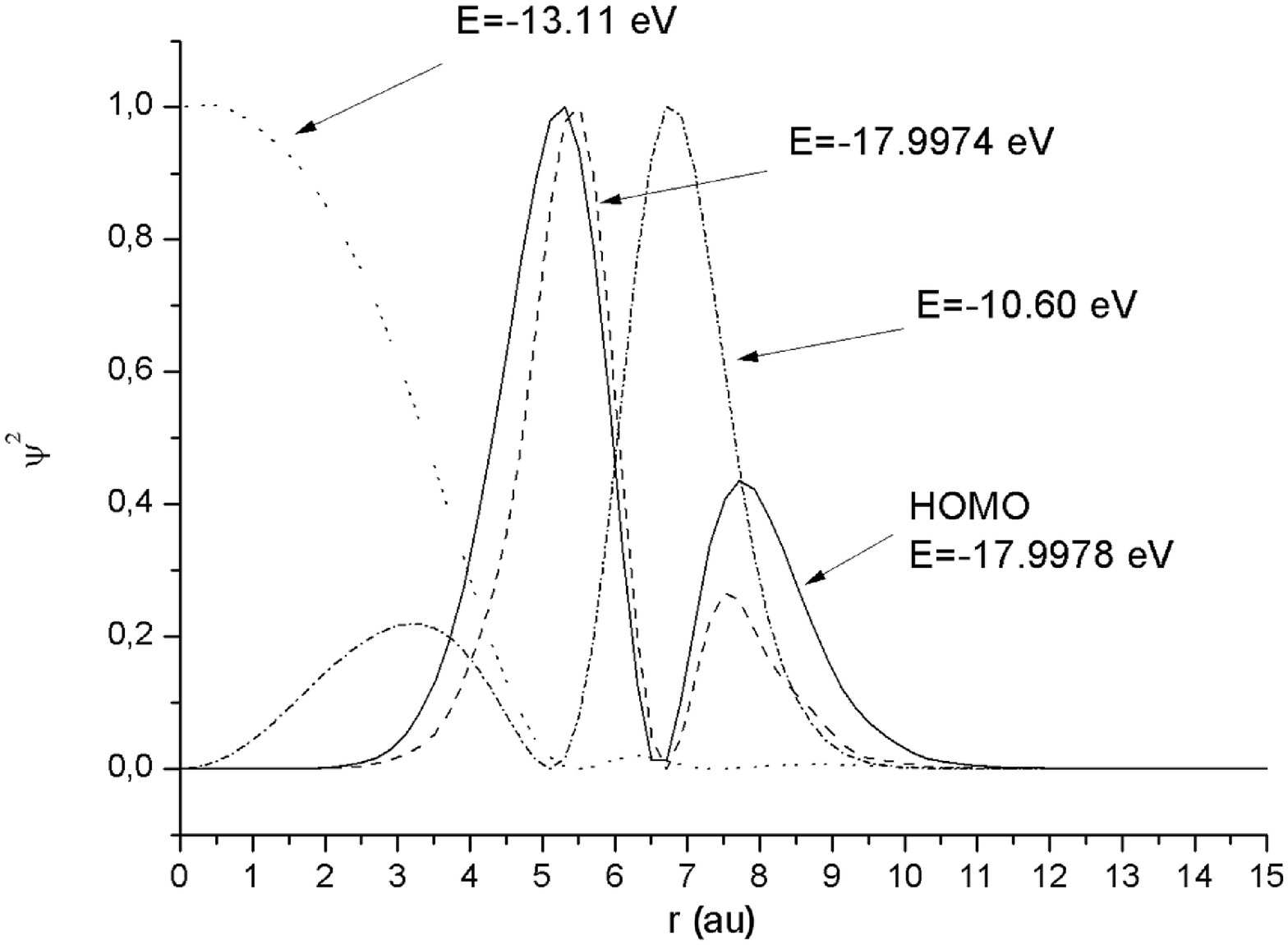}
	\caption{Wave functions obtained with the help of the DFT computations for a charged $C_{60}$ fullerene with $Z = 5e$: both the surface-localized (solid and broken curves) and the volume-localized (two dot-style curves) states are found. For convenience of comparison, the wave functions are shown by normalizing their maximal values to unit.}
	\label{dft_states}
  \end{center}
\end{figure}

The spherically symmetric analytical model potential (\ref{7}) can be compared with the result derived from DFT computations. Fig.~\ref{dft} shows potential's behavior along the radial direction through carbon's atom (``atom'' curve) and through the center between neighboring atoms (``middle'' curve), respectively; averaging over the angles yields a better agreement. Furthermore, one can directly confirm the existence of the volume-localized states by making use of the Quantum Espresso package; the results of the corresponding DFT computations are presented in Fig.~\ref{dft_states}, and it is satisfactory to see the consistency with the results in Fig.~\ref{vol} obtained in the framework of the model potentials approach. 

\section{Onion-like fullerene}

Onion (or onion-like) structures represent a highly interesting class of carbon systems which are obtained when fullerenes are concentrically enclosed one into another to form a double-, triple-, or in general a multi-layered object. Each (quasi-)spherical layer is a fullerene ($C_{60}$, $C_{240}$, $C_{540}$, $C_{960}$, $\cdots$), with the separation between shells equal to $0.335\,$nm which is slightly larger than the distance between the planar layers in graphite crystals. The outer diameter of a typical 5-15 layered onion ranges between 4 to 10 nm, with the radius of the innermost layer equal to $R_f$, however smaller and much larger onion-like structures are also observed. Following the first observation \cite{Iijima:1980} of spherical multi-layered structures, the concept of ``carbon onion'' was coined in 1992 when the formation of the onion-like spherical particles was demonstrated by heating of nanotubes with an electron beam \cite{Ugarte:1992,Ugarte:1993,Ugarte:1995}. Since then the physical and chemical characteristics of carbon onions was analyzed in numerous theoretical and experimental studies \cite{Zwanger:1996,Tomita:1999,Okotrub:2001,Blank:2018}; see the reviews \cite{Shenderova:2002,Xu:2008,Butenko:2011,Chang:2011,He:2011,Bartelmess:2014,Georgakilas:2015,Mykhailiv:2017,Butenko:2017} for the further information on the production, geometrical, physical and chemical properties, and applications of onion-like carbon structures.

In the context of the current investigation of the energy spectrum of charged carbon complexes, the onion-like structures are qualitatively different from the usual fullerenes in the sense that, in contrast to the latter case when the electric charge is smeared only over the surface of fullerene's hollow sphere, in the former case the electric charge is distributed in the volume of an onion sphere on its many internal layers. 

Accordingly, in a simplest model for the study of discrete volume levels of electrons in multi-layer onion-like charged fullerenes (somewhat similarly to the simplest model (\ref{2}) and (\ref{3}) of a rectangular spherical well for a fullerene), one can look for analytic estimates by assuming a homogeneous density when the charges of consecutive layers of the onion structure are proportional to the cube of the layer radius. In this case, the potential energy of an electron in an electrostatic field is as follows:
\begin{equation}\label{8}
  U(r) = -\,{\frac {Ze}{4\pi\varepsilon_0}}\times\left \{ \begin{array}{cc}
  {\displaystyle {\frac {1}{2R}}\left(3 - \frac{r^2}{R^2}\right) },  &  r \leq R , \\
   & \\
  {\displaystyle {\frac 1r} },  &  r > R . \\
\end{array}\right . 
\end{equation} 
Here $Z$ is the total positive charge of an onion structure, and $R = R_{\rm on}$ is its outer radius. As a first step to understand the spectrum structure, we approximate the potential by extending the piece inside the sphere $r \leq R$ to all values of the radius:
\begin{eqnarray}
U(r) = -\,U_0 + {\frac 12}\,m_e\omega^2r^2,\label{8a}
\end{eqnarray}
where we denoted 
\begin{equation}\label{omega}
U_0 = {\frac 32}{\frac{Ze}{4\pi\varepsilon_0R_{\rm on}}},\qquad
\omega = \sqrt{\frac{Ze}{4\pi\varepsilon_0m_eR_{\rm on}^3}}.
\end{equation} 

The maximal specific charge $Z/N_{\rm tot}$ of onion-like structures (where $N_{\rm tot}$ is the total number of atoms) before their decay would be smaller than that for $C_ {60}$, but the absolute value of the charge can be much larger. Accordingly, the depth of the potential well $U_0$ of an electron in the field of a positively charged onion structure then can reach the values of order of 100 eV, thereby increasing the significance of the volume-localized quantum states.

For the approximate potential (\ref{8}), one can evaluate the energy spectrum analytically by making use of the well-known solution of the Schr\"odinger equation for the spherical oscillator \cite{Landau}. For energy levels we find
\begin{eqnarray}\label{9}
E &=& E_0 + \hbar\omega\left(2i + l - 2\right),\\
E_0 &=& {\frac 32}\sqrt{\frac{Ze}{4\pi\varepsilon_0R_{\rm on}}}\left(\sqrt{\frac{\hbar^2}
{m_eR_{\rm on}^2}} - \sqrt{\frac{Ze}{4\pi\varepsilon_0R_{\rm on}}}\right),\label{E0}
\end{eqnarray}
whereas the wave functions of the corresponding stationary states are 
\begin{eqnarray}
\psi_{nlm} = \mathrm{const}\,r^l \exp \left(-\,\frac{\lambda r^2}{2} \right)
Y_{lm} (\theta, \varphi)\,\times \nonumber\\
\times\,{}_1\!F_1\Bigl(1 - i, l + \frac{3}{2}, \lambda r^2\Bigr),\label{10}
\end{eqnarray} 
where $_1\!F_1$ is the degenerate hypergeometric function,
\begin{equation}\label{11}
\lambda = {\frac {m_e\omega}{\hbar}} = \sqrt{\frac{Ze\,m_e}{4\pi\varepsilon_0\hbar^2R_{\rm on}^3}},
\end{equation}
the radial quantum number $i = 1, 2, \cdots$, the angular quantum number $l = 0, 1, 2, \cdots$, and $m = 0, \pm 1, \cdots, \pm l$.

\begin{figure}
  \begin{center}
    \includegraphics[width=1.\linewidth] {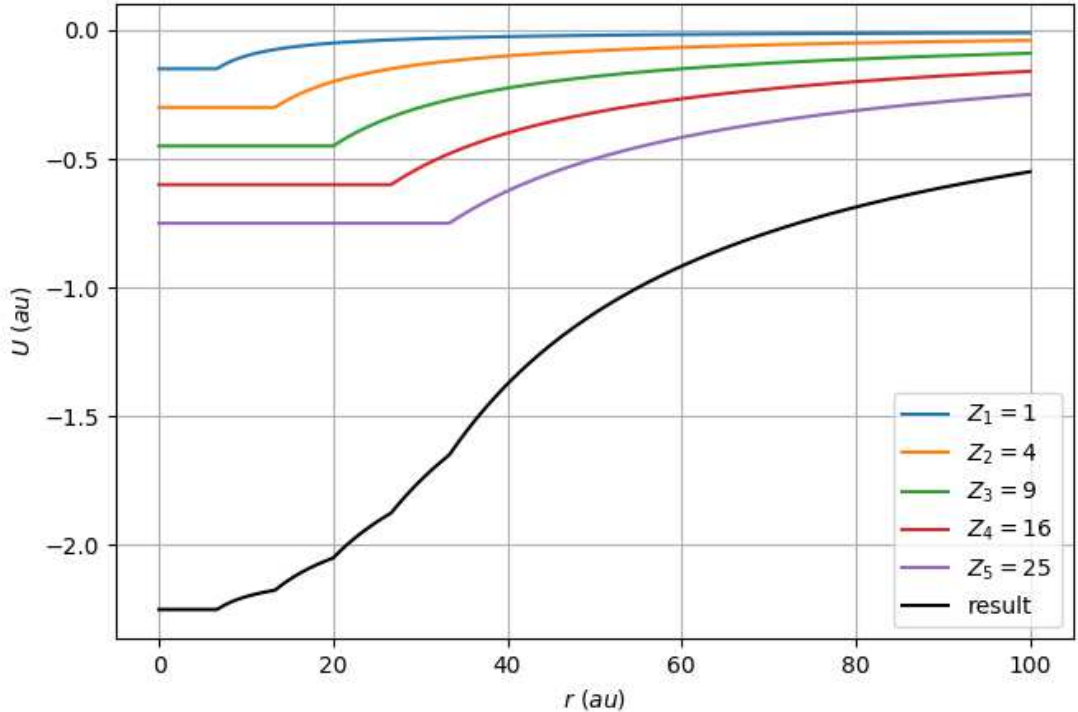}
\caption{Model of charged onion-like fullerene: contributions of layers and the resulting potential (\ref{13}) with regular charge distribution (\ref{choice1}).}
	\label{f-22}
  \end{center}
\end{figure}

\begin{figure}
  \begin{center}
    \includegraphics[width=1.\linewidth] {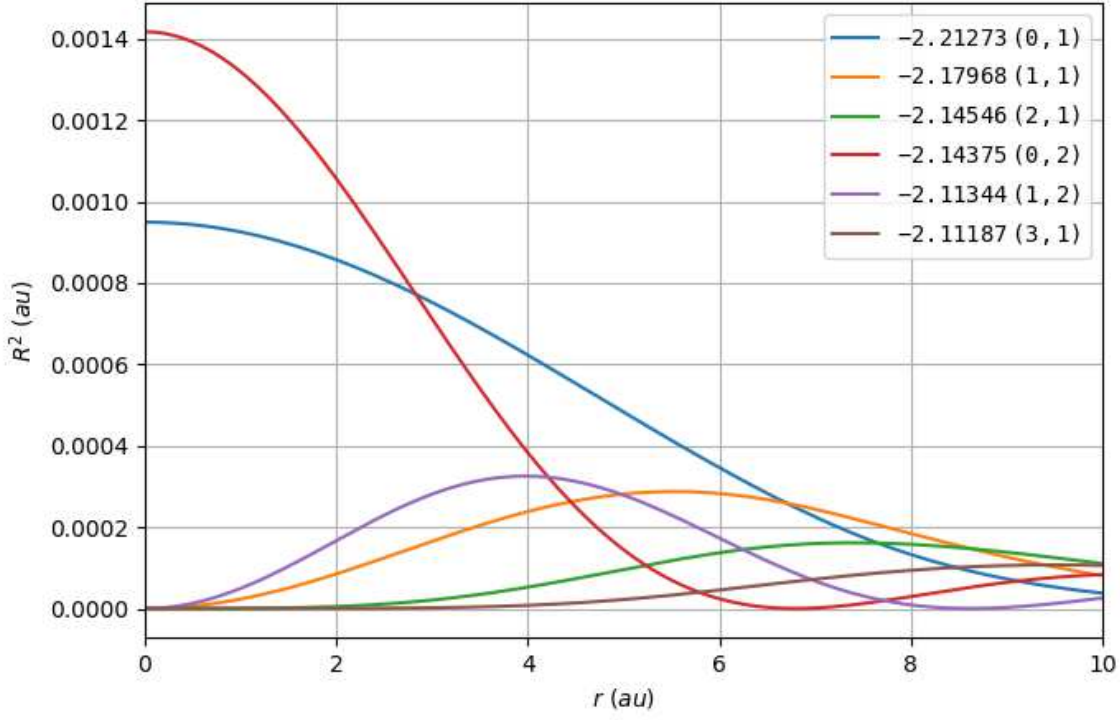}
\caption{Wave functions for charged onion-like fullerene model potential (\ref{13}) with regular charge distribution (\ref{choice1}).}
	\label{f-23}
  \end{center}
\end{figure}

\begin{figure}
  \begin{center}
    \includegraphics[width=1.\linewidth] {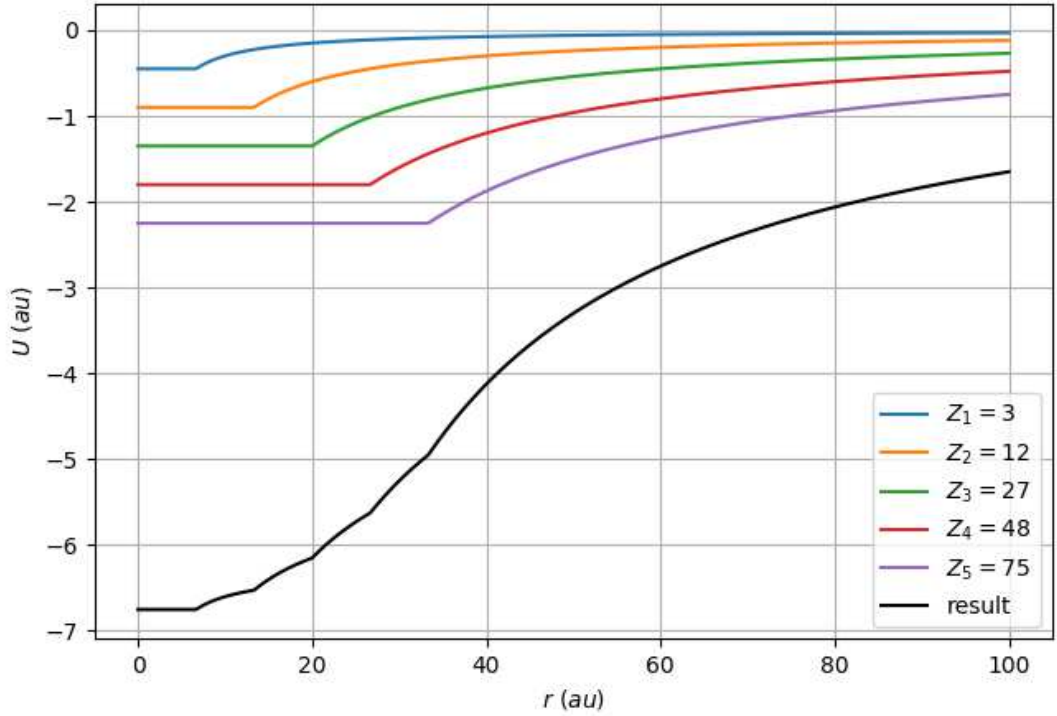}
\caption{Model of charged onion-like fullerene: contributions of layers and the resulting potential (\ref{13}) with irregular charge distribution (\ref{choice2}).}
	\label{f-26}
  \end{center}
\end{figure}

As a particular application, let us consider a model of a 5-layer charged onion fullerene with $Z = 55e$ and the size $R_{\rm on} = 5R_f$. The total charge arises from the assumption of a homogeneous distribution of the electric charge on the inner layers proportionally to the third power radius of the layer. The corresponding energy levels (\ref{9}) for such onion model are presented in Table~\ref{t-on}. The well-known degeneracy properties of an oscillator spectrum are manifest. It is instructive to compare these analytic estimates with the results of a numeric integration of the Schr\"odinger equation for a five-layered onion structure which are given in Fig.~\ref{f-22}-\ref{f-27} and in Tables~\ref{t-17}-\ref{t-19} for different values of the positive charge of the layers. 

\begin{table}
\caption{Electron energy levels (\ref{9}) for an onion-like fullerene potential (\ref{8}) with $Z = 55$ and $R_{\rm on} = 33.135 a_0$. [Notation: $n$ -- level number, $l$ -- angular quantum number, $i$ -- radial quantum number].}\label{t-on}
  \begin{tabular}{ccccc|ccccc}
  \hline
   $n$  & $l$ & $i$ & $E$ (au)  & & & $n$  & $l$ & $i$ & $E$ (au) \\
  \hline
  1    &     0     &     1    &  $-2.48981$   & & &
  7    &     2     &     2    &  $-2.45093$    \\  
  2    &     1     &     1    &  $-2.41205$  & & &
  8    &     0     &     3    &  $-2.37316$    \\ 
  3    &     2     &     1    &  $-2.33428$   & & &
  9    &     4     &     1    &  $-2.29540$    \\
  4    &     0     &     2    &  $-2.25652$   & & &
 10    &     3     &     2    &  $-2.21763$    \\
  5    &     3     &     1    &  $-2.17875$   & & &
 11    &     1     &     3    &  $-2.13987$    \\ 
  6    &     1     &     2    &  $-2.10099$   & & &
 12    &     5     &     1    &  $-2.06211$    \\  
  \hline
  \end{tabular} 
\end{table} 

Theoretical analysis of the structure of discrete levels for electrons in the field of positively charged single- and multi-layer fullerenes suggests the existence of a system of quantum transitions with emission of photons within a wide range of energies that depend on the magnitude of the charge and the lifetime of the levels in the range from 21 ns to 21 fs. 

In order to make estimates, let us recall that the characteristic lifetime for the spontaneous dipole transition is determined by the well-known expression for the photon emission rate 
\begin{equation}\label{12}
 P_{fi} = {\frac{\omega^3}{3 \pi\varepsilon_0\hbar c^3}} |d_{fi}|^2 ,
\end{equation} 
where $\omega$ is the emission frequency and $d_{fi}$ is the matrix element of the dipole transition from an initial ({\it i}) to a final ({\it f}) state. One can roughly estimate the order of magnitude of $d_{fi}$ for transitions between the volume-localized discrete levels as $eR_f$ for charged fullerenes and as $eR_{\rm on}$ for onion-like structures. These processes may occur during pulse charging of fullerenes or in the process of irradiation of already charged fullerenes by a flow of electrons and ions. Experimental confirmation of the existence of the volume-localized discrete levels is of great interest for the experimental research and practical problems including a development of the new sources of coherent radiation in a wide range of wavelengths. 

Simple numeric estimates are straightforward. Substituting the approximation $|d_{fi}| \sim eR_f$, we recast (\ref{12}) into
\begin{eqnarray}
P_{fi} = {\frac {4(6.627)^2}{3\alpha}}{\frac{c}{\lambdabar_e}}
\left({\frac {\Delta E}{m_ec^2}}\right)^3.\label{T}
\end{eqnarray} 
Here we assumed $R_f = 6.627 a_0$ for the radius of the fullerene. As a result, we find for different energy transitions: $P_{fi}(\Delta E = 1\,{\rm eV}) = 4.67 \times 10^7\,$s$^{-1}$, $P_{fi}(\Delta E = 10\,{\rm eV}) = 4.67 \times 10^{10}\,$s$^{-1}$, $P_{fi}(\Delta E = 100\,{\rm eV}) = 4.67 \times 10^{13}\,$s$^{-1}$. More accurate estimates for transition rates in the system of discrete volume-localized levels of an electron in the field of a positively charged fullerene can be derived from the evaluation of the value $<\!\!er\!\!>$ by making use of the wave functions of the spherical oscillator (\ref{10}). The characteristic length of the exponential decay part of the wave function is then found as $1/\sqrt{\lambda} = \left(R_{\rm on}^3a_0e/Z\right)^{1/4}$. 

To complete the discussion of the onion-like structures, it is important to analyze the possible values of $Z$ for onion fullerenes. The corresponding experimental and theoretical results on the limiting values of the positive electric charge for ionized onion-like fullerenes are absent. Nevertheless, one can make some simple estimates for the highest value of the charge by evaluating the critical value of the field strength on the outer spherical layer of an onion fullerene. 

In order to do this, we start with the case of an ordinary fullerene and notice that, for a positive charge $Z$ on it, the value of the electric field strength reads ${\mathcal E} = {\frac {Z}{4\pi\varepsilon_0R_f^2}}$, evaluated at the spherical surface of a fullerene, before it becomes unstable due to the field ion emission. Taking $R_f = 6.627 a_0$ and $Z = 12e$ in the fullerene $C_{60}$, we find for the field strength ${\mathcal E} = 1.38 \times 10^{11}\,$V/m. This is smaller than the critical value (evaporation field) for the carbon ${\mathcal E}_{\rm max} = 1.48 \times 10^{11}\,$V/m, above which the ion field emission starts \cite{Muller:1960,Forbes:2003}. One can reasonably assume that this field value should not be exceeded also for the onion-like structures. 

We thus can formulate a simple criterion ${\mathcal E} \apprle {\mathcal E}_{\rm max}$ for the stability of a charged multi-layer onion fullerene, with the help of which one can derive a rough estimate of the corresponding maximal possible total positive charge $Z = Ne$. In particular, applying this scheme to the 5-layer onion structure with $R_{\rm on} = 5R_f$ and $N = \sum_i N_i = 55$, we find ${\mathcal E} = 2.64\times 10^{10}\,$V/m which is well below the threshold value. In a similar way, one can evaluate the limiting charge for onion-like structures with an arbitrary number of layers. 

Obviously, such a semi-empirical estimate is very approximate and needs to be further refined on the basis of the microscopic calculations or the experimental measurements. 

\section{Numeric analysis of onion-like fullerene model}

The analytic model which we considered in the previous section, treats a charged onion fullerene as a homogeneously charged solid sphere. In other words, in such an approach the internal structure is not taken into account. In order to improve the picture, we replace the model (\ref{8}) with a potential that explicitly describes an onion fullerene as a multi-layer structure with the concentric charged spheres enclosed one into another. For concreteness, let us consider a simple model of an onion-like fullerene with five layers. We construct the corresponding potential in the form of a superposition of the five Coulomb-like contributions (\ref{1}) coming from each of the charged spherical shell  
\begin{equation}\label{13}
 U(r) = \sum_{k=1}^{5} U_k(r)\,,
\end{equation} 
where
\begin{equation}\label{Uk}
 U_k(r) = -\,{\frac {N_ke^2}{4\pi\varepsilon_0}}\times\left\{\begin{array}{cc}
  {\displaystyle {\frac {5}{kR_{\rm on}}} },  &  r \leq {\frac k5}R_{\rm on} , \\
   & \\
  {\displaystyle {\frac {1}{r}} },  &  r > {\frac k5}R_{\rm on} . \\
\end{array}\right . 
\end{equation}
Here $Z_k = N_ke$ is the charge on a layer number $k = 1,2, ..., 5$. The choice of the values of the electric charge on shells is a highly nontrivial issue which basically should be determined by the physical procedure used to charge the layers. Below, we analyze the two options. 

\begin{table}
\caption{Electron energy levels for charged onion-like fullerene model potential (\ref{13}) with regular charge distribution (\ref{choice1}). [Notation: $n$ -- level number, $l$ -- angular quantum number, $i$ -- radial quantum number].}\label{t-17}
  \begin{tabular}{ccccc|ccccc}
  \hline
   $n$  & $l$ & $i$ & $E$ (au)  & & & $n$  & $l$ & $i$ & $E$ (au) \\
  \hline
   1    &     0     &     1    &  $-2.21273$   & & &
   7    &     2     &     2    &  $-2.08023$  \\
   2    &     1     &     1    &  $-2.17968$   & & &
   8    &     4     &     1    &  $-2.07870$  \\
   3    &     2     &     1    &  $-2.14546$   & & &
   9    &     0     &     3    &  $-2.07677$  \\
   4    &     0     &     2    &  $-2.14375$   & & &
  10    &     5     &     1    &  $-2.04537$  \\
   5    &     1     &     2    &  $-2.11344$   & & &
  11    &     3     &     2    &  $-2.04468$  \\
   6    &     3     &     1    &  $-2.11187$   & & &
  12    &     1     &     3    &  $-2.04256$  \\
  \hline
  \end{tabular} 
\end{table} 

\subsection{Regular charge distribution}

Suppose, one can arrange a charged onion structure in such a way that it closely reproduces a model of a uniformly charged solid sphere. We will call this a regular charge distribution. In this case, the values of the charge on the shells are proportional to the square of layer's radius:  
\begin{equation}\label{choice1}
 N_1 = 1,\ N_2 = 4,\ N_3 = 9,\ N_4 = 16,\ N_5 = 25\,.
\end{equation}
The resulting potential $U(r)$ and its constituents (\ref{10}) are shown in Fig.~\ref{f-22}. The spectrum is presented on Fig.~\ref{f-23} and Table~\ref{t-17}. The numeric computations should be compared to the estimates obtained on the basis of analytic model which we studied in the previous section. The data in Table~\ref{t-on} and Table~\ref{t-17} are in a very good agreement. 

Although the total charge of such onion fullerene $Z = 55e$ is large, the resulting specific charge (per total number of atoms) is actually smaller than that of the maximally charged $C_{60}$. Moreover, even with such a large total charge, this onion fullerene model satisfies the simple criterion of stability since the electric field on its boundary does not exceed the critical value for carbon.

\subsection{Irregular charge distribution}

\begin{figure}
  \begin{center}
    \includegraphics[width=1.\linewidth] {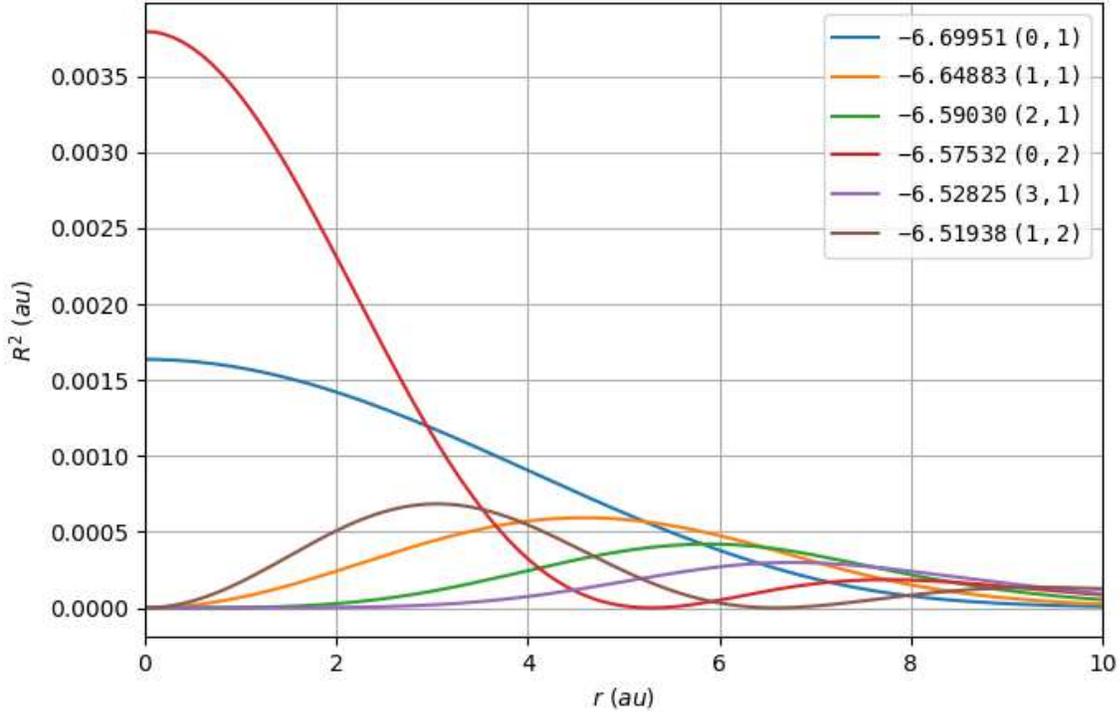}
\caption{Wave functions for charged onion-like fullerene model potential (\ref{13}) with irregular charge distribution (\ref{choice2}).}
	\label{f-27}
  \end{center}
\end{figure}

Obviously, it is a highly nontrivial technical problem to produce a regularly charged onion fullerene. It is more likely that in practice, in the course of an actual experimental laboratory setup, an onion structure can be only charged in an irregular way. In that case, one should expect to find that the charges on each shell would have more or less arbitrary values. In order to analyze the corresponding spectrum numerically, let us choose
\begin{equation}\label{choice2}
 N_1 = 3,\ N_2 = 12,\ N_3 = 27,\ N_4 = 48,\ N_5 = 75\,.
\end{equation}
The resulting potential $U(r)$ and its constituents (\ref{10}) are shown in Fig.~\ref{f-26}. The spectrum is presented on Fig.~\ref{f-27}, and the first energy eigenvalues for the irregular choice are listed in Table~\ref{t-19}. 

For such an irregularly charged onion fullerene, the total charge is $Z = 165e$, and the value of the electric field strength on the outer boundary ${\mathcal E} = 7.6\times 10^{10}\,$V/m is well below the stability limit. As a certain self-consistency test, one can verify that, keeping the same total charge, the mostly positive area is at the surface (the outmost shell) of an onion-like fullerene which results in more shallow well with a smaller depth of the potential at the centre. 

\begin{table}
\caption{Electron energy levels for charged onion-like fullerene model potential (\ref{13}) with irregular charge distribution (\ref{choice2}). [Notation: $n$ -- level number, $l$ -- angular quantum number, $i$ -- radial quantum number].}\label{t-19}
  \begin{tabular}{ccccc|ccccc}
  \hline
   $n$  & $l$ & $i$ & $E$ (au)  & & & $n$  & $l$ & $i$ & $E$ (au) \\
  \hline
  1    &     0     &     1    &  $-6.69951$   & & &
  7    &     2     &     2    &  $-6.47005$    \\  
  2    &     1     &     1    &  $-6.64883$  & & &
  8    &     0     &     3    &  $-6.46943$    \\ 
  3    &     2     &     1    &  $-6.59030$   & & &
  9    &     4     &     1    &  $-6.46691$    \\
  4    &     0     &     2    &  $-6.57532$   & & &
 10    &     3     &     2    &  $-6.41962$    \\
  5    &     3     &     1    &  $-6.52825$   & & &
 11    &     1     &     3    &  $-6.41376$    \\ 
  6    &     1     &     2    &  $-6.51938$   & & &
 12    &     5     &     1    &  $-6.40848$    \\  
  \hline
  \end{tabular} 
\end{table}

\section{Discussion and conclusion}

In the framework of a simple physical model, we demonstrate the existence of a system of discrete short-lifetime quantum levels for electrons in the potential well of the self-consistent Coulomb field of charged fullerenes and onion-like structures. For electrons, in the case of positively charged fullerenes and onion-like structures, the energy of the volume-localized levels ranges from 1~eV to 100 eV.

We use an idealized spherically symmetric model potentials in our study. As one knows, geometrically a $C_{60}$ fullerene is a truncated icosahedron with carbon atoms located in its 60 equivalent vertices. Such a geometrical shape is very close to a sphere and it is possible to use the spherical harmonics to describe the electronic structure of a neutral $C_{60}$ in terms of free electrons on a sphere of radius $R$. The corresponding symmetries of the icosahedron group are $\Gamma_{60} = A_g + T_{1g} + 2T_{1u} + T_{2g} + 2T_{2u} + 2G_g + 2G_u + 3H_g + 2H_u$, and with the help of a simple H\"uckel scheme one can calculate the eigenvalues and degeneracies of this system \cite{Fowler:1986}. (In fact, the first such calculation by Bochvar and Gal'pern \cite{Bochvar:1973} predicted the existence of stable carbon structures with the icosahedron geometry a decade before the experimental discovery of fullerenes). The H\"uckel scheme calculation is perfectly consistent with the density functional theory computation \cite{Saito:1992} that explicitly demonstrates the correspondence of $A_g$, $T_{1u}$, $H_g$, $T_{2u} + G_u$ to $l = 0,1,2,3$ ($1s, 1p, 1d, 1f$) states, respectively; see also \cite{Verkhovtsev:2013} for the higher quantum angular numbers up to $l = 9$.

It is worthwhile to mention that the electronic structure of fullerenes (and, in general, interaction of electrons and photons with fullerenes and fullerene-like systems) is widely studied by means of the simple model potentials. In particular, along with the square-well type potentials one considers the Dirac bubble and Gaussian-type potentials, as well as models the fullerene cage by a spherical jellium shell \cite{Jaskolski:1996,Lohr:1992,Puska:1993,Amusia:1998,Connerade:1999a,Connerade:1999b,Schrange:2016,Felfli:2018}. The previous discussions were mostly focused on the neutral systems, whereas here we have analysed the case of positively charged fullerenes and onion-like structures. 

The results obtained provide a consistent qualitative picture both for the charged fullerenes and for the onion-like structures. In order to refine our very approximate findings, one certainly needs a further investigation on the basis of the microscopic calculations, as well as the experimental measurements.

An experimental confirmation of the existence of the volume-localized discrete levels would be of great interest for the experimental research and practical problems including a development of the new sources of coherent radiation in a wide range of wavelengths. 


\begin{acknowledgments}

We thank P.~S. Kondratenko and participants of the seminar of the Theoretical Physics Laboratory, Institute for Nuclear Safety (IBRAE) for the fruitful and stimulating discussions. We are grateful to A.~V. Osadchy for the help with making the DFT computations and producing the corresponding Figures \ref{dft} and \ref{dft_states}.

\end{acknowledgments}


\begin{thebibliography}{99}

\bibitem{Kroto:1985}
W.~H. Kroto, J.~R. Heath, S.~C. O'Brien, R.~F. Curl, and R.~E. Smalley,
$C_{60}$: Buckminsterfullerene,
Nature {\bf 318}, 162 (1985). 

\bibitem{Kraetschmer:1990}
W. Kr\"atschmer, L.~D. Lamb, K. Fostiropoulos, and D.~R.~Huffman,
Solid $C_{60}$: a New Form of Carbon,
Nature {\bf 347}, 354 (1990). 

\bibitem{Eletski:1995}
A.~V. Eletskii and B.~M. Smirnov, Fullerenes and carbon structures,
Phys. Usp. {\bf 38}, 935 (1995). 

\bibitem{Hirsch:2005}
A. Hirsch and M. Brettreich, {\it Fullerenes: Chemistry and Reactions}
(Wiley-VCH Verlag: Weinheim, 2005).

\bibitem{Sattler:2010}
K.~D. Sattler, ed., {\it Handbook of nanophysics. Clusters and fullerenes}
(CRC Press: Boca Raton, 2011).

\bibitem{Gogotsi:2010}
Yu. Gogotsi, ed., {\it Nanomaterials Handbook}, 2nd ed.
(CRC Press: Boca Raton, 2017).

\bibitem{Campbell}
E.~E.~B. Campbell, {\it Fullerene Collision Reactions}
(Kluwer Academic Publ.: N.Y., 2004). 

\bibitem{Radi:1990}
P.~P. Radi, M.-T. Hsu, M.~E. Rincon, P.~R. Kemper, and M.~T. Bowers, 
On the structure, reactivity and relative stability of the
large carbon cluster ions $C^+_{62}$, $C^+_{60}$ and $C^+_{58}$,
Chem. Phys. Lett. {\bf 174}, 223 (1990). 

\bibitem{Troullier:1992}
N. Troullier and J.~L. Martins, 
Structural and electronic properties of $C_{60}$,
Phys. Rev. A {\bf 46}, 1754 (1992). 

\bibitem{Yabana:1993}
K. Yabana and G.~F. Bertsch,
Electronic structure of $C_{60}$ in a spherical basis,
Physica Scr. {\bf 48}, 633 (1993). 

\bibitem{Brenac:1999}
A. Brenac, F. Chandezon, H. Lebius, A. Pesnelle, S.~Tomita, and B.~A. Huber,
Multifragmentation of Highly Charged C60 Ions: Charge States and Fragment Energies,
Physica Scr. {\bf T80B}, 195 (1999). 

\bibitem{Bhardwaj:2003}
V.~R. Bhardwaj, P.~B. Corkum, and D.~M. Rayner,
Internal Laser-Induced Dipole Force at Work in $C_{60}$ Molecule,
Phys. Rev. Lett. {\bf 91}, 203004 (2003). 

\bibitem{Jensen:2004}
J. Jensen, H. Zettergren, H.~T. Schmidt, H. Cederquist, S. Tomita, S. B. Nielsen,
J. Rangama, P. Hvelplund, B.~Manil, and B.~A. Huber,
Ionization of $C_{70}$ and $C_{60}$ molecules by slow highly charged ions: A comparison,
Phys. Rev. A {\bf 69}, 053203 (2004). 

\bibitem{Zettergren:2002}
H. Zettergren, H.~T. Schmidt, H. Cederquist, J. Jensen, S. Tomita, P. Hvelplund,
H. Lebius, and B.~A. Huber,
Static over-the-barrier model for electron transfer between metallic spherical objects,
Phys. Rev. A {\bf 66}, 032710 (2002). 

\bibitem{Zettergren:2004}
H. Zettergren, J. Jensen, H. T. Schmidt, and H. Cederquist,
Electrostatic model calculations of fission barriers for fullerene ions,
Eur. Phys. J. D {\bf 29}, 63 (2004). 

\bibitem{Diaz:2005a}
S. D\'iaz-Tendero, M. Alcam\'i, and F. Mart\'in, 
Coulomb Stability Limit of Highly Charged $C^{q+}_{60}$ Fullerenes,
Phys. Rev. Lett. {\bf 95}, 013401 (2005). 

\bibitem{Diaz:2005b}
S. D\'iaz-Tendero, M. Alcam\'i, and F. Mart\'in, 
Structure and electronic properties of highly charged $C_{60}$ and $C_{58}$ fullerenes,
J. Chem. Phys. {\bf 123}, 184306 (2005). 

\bibitem{Sahnoun:2006a}
R. Sahnoun, K. Nakai, Y. Sato, H. Kono, Y. Fujimura, and M. Tanaka,
Theoretical investigation of the stability of highly charged $C_{60}$
molecules produced with intense near-infrared laser pulses,
J. Chem. Phys. {\bf 125}, 184306 (2006). 

\bibitem{Sahnoun:2006b}
R. Sahnoun, K. Nakai, Y. Sato, H. Kono, Y. Fujimura, and M. Tanaka,
Stability limit of highly charged $C_{60}$ cations produced with an
intense long-wavelength laser pulse: Calculation of electronic
structures by DFT and wavepacket simulation,
Chem. Phys. Lett. {\bf 430}, 167 (2006). 

\bibitem{Iroshnikov:2006}
G.~S. Iroshnikov, 
Calculation of the cross section for charge transfer in
fullerene-fullerene collisions,
JETP {\bf 103}, 707 (2006). 

\bibitem{Iroshnikov:2007}
G.~S. Iroshnikov, 
Calculation of the cross section for charge transfer in
$C^+_{70} + C_{60}$ fullerene-fullerene collisions,
JETP {\bf 105}, 706 (2007). 

\bibitem{Kono:2007}
H. Kono, K. Nakai, and N. Niitsu,
Ab initio molecular dynamics of highly charged fullerene cations
in intense near-infrared laser fields,
Proc. of SPIE {\bf 6726}, 672606 (2007). 
  
\bibitem{Madjet:2008}
M.~E. Madjet, H.~S. Chakraborty, J.~M. Rost, and S.~T.~Manson,
Photoionization of $C_{60}$: a model study,
J. Phys. B: At. Mol. Opt. Phys. {\bf 41}, 105101 (2008). 

\bibitem{Nascimento:2011}
E.~M. Nascimento, F.~V. Prudente, M.~N. Guimar\~{a}es, and A.~M. Maniero,
A study of the electron structure of endohedrally confined atoms
using a model potential,
J. Phys. B: At. Mol. Opt. Phys. {\bf 44}, 015003 (2011). 

\bibitem{Polozkov:2005}
R.~G. Polozkov, V.~K. Ivanov, and A.~V. Solov'yov,
Photoionization of the fullerene ion $C^+_{60}$,
J. Phys. B: At. Mol. Opt. Phys. {\bf 38}, 4341 (2005). 

\bibitem{Polozkov:2013}
R.~G. Polozkov, V.~K. Ivanov, A.~V. Verkhovtsev, A.~V.~Korol, and A.~V. Solov'yov,
New applications of the jellium model for the study of atomic clusters,
J. Phys.: Conf. Ser. {\bf 438}, 012009 (2013). 
  
\bibitem{Luders:2009}
M. L\"uders, A. Bordoni, N. Manini, A. Dal Corso, M.~Fabrizio, and E. Tosatti,
Coulomb couplings in positively charged fullerene,
Phil. Mag. B {\bf 82}, 1611 (2009). 

\bibitem{Liu:2018}
D. Liu, N. Iwahara, and L.~F. Chibotaru, 
Dynamical Jahn-Teller effect of fullerene anions,
Phys. Rev. B {\bf 97}, 115412 (2018). 

\bibitem{Wang:2011}
Y. Wang, M. Alcam\'i, and F. Mart\'in, 
Stability of charged fullerenes,
in: {\it Handbook of nanophysics. Clusters and fullerenes}, Ed. K.~D. Sattler
(CRC Press: Boca Raton, 2011) 25. 

\bibitem{preprint}
R.V. Arutyunyan,
Theoretical investigation of electronic properties of highly charged fullerenes,
Preprint No. IBRAE-2018-08 (Inst. Nucl. Safety IBRAE, Moscow, 2018) 16 p. 

\bibitem{Landau}
L.~D. Landau and E.~M. Lifshitz, {\it Quantum mechanics. Non-relativistic theory},
3rd ed. (Pergamon Press: Oxford, 1977).   

\bibitem{Rubio:1994}
A. Rubio, J.~A.~Alonso, J.~M.~L\'opez, and M.~J.~Stott,
Collective electronic excitations in metal coated $C_{60}$, 
Phys. Rev. B {\bf 49}, 17397 (1994). 

\bibitem{Ivanov:2001}
V.~K. Ivanov, G.~Yu. Kashenock, R.~G. Polozkov, and A.~V.~Solov'yov,
Photoionization cross sections of the fullerenes $C_{20}$ and
$C_{60}$ calculated in a simple spherical model,
J. Phys. B: At. Mol. Opt. Phys. {\bf 34}, L669 (2001). 

\bibitem{Ivanov:2003}
V.~K. Ivanov, G.~Yu. Kashenock, R.~G. Polozkov, and A.~V.~Solov'yov,
Method for calculating photoionization cross sections of fullerenes
in the local density and random phase approximations,
JETP {\bf 96}, 658 (2003). 

\bibitem{Belyaev:2009}
A.~K. Belyaev, A.~S. Tiukanov, A.~I. Toropkin, V.~K.~Ivanov, R.~G. Polozkov,
and A.~V. Solov'yov,
Photoabsorption of the fullerene $C_{60}$ and its positive ions,
Physica Scr. {\bf 80}, 048121 (2009). 

\bibitem{Verkhovtsev:2012}
 A.~V.~Verkhovtsev, R.~G. Polozkov, V.~K.~Ivanov, A.~V.~Korol, and A.~V.~Solov'yov,
Hybridization-related correction to the jellium model for fullerenes,
J. Phys. B: At. Mol. Opt. Phys. {\bf 45}, 215101 (2012). 

\bibitem{Baltenkov:2015}
A.~S. Baltenkov, S.~T. Manson, and A.~Z. Msezane,
Jellium model potentials for the $C_{60}$ molecule and the
photoionization of endohedral atoms, $A$@$C_{60}$,
J. Phys. B: At. Mol. Opt. Phys. {\bf 48}, 185103 (2015). 

\bibitem{Iijima:1980}
S. Iijima,
Direct observation of the tetrahedral bonding in graphitized carbon black by HREM,
J. Crystal Growth {\bf 50}, 675 (1980). 

\bibitem{Ugarte:1992}
D. Ugarte, 
Curling and closure of graphitic networks under electron-beam irradiation,
Nature {\bf 359}, 707 (1992). 

\bibitem{Ugarte:1993}
D. Ugarte, How to fill or empty a graphitic onion,
Chem. Phys. Lett. {\bf 209}, 99 (1993). 

\bibitem{Ugarte:1995}
D. Ugarte, Onion-like graphitic particles,
Carbon {\bf 33}, 989 (1995). 

\bibitem{Zwanger:1996}
M.~S. Zwanger, F. Banhart, and A. Seeger,
Formation and decay of spherical concentric-shell carbon clusters,
J. Crystal Growth {\bf 163}, 445 (1996). 

\bibitem{Tomita:1999}
S. Tomita, M. Fujii, S. Hayashi, and K. Yamamoto,
Electron energy-loss spectroscopy of carbon onions,  
Chem. Phys. Lett. {\bf 305} (1999) 225. 

\bibitem{Okotrub:2001}
A.~V. Okotrub, L.~G. Bulusheva, A.~I. Romanenko, V.~L.~Kuznetsov, Yu.~V. Butenko,
C. Dong, Y. Ni, and M.~I.~Heggie,
Probing the electronic state of onion-like carbon,
AIP Conference Proceedings {\bf 591}, 349 (2001). 

\bibitem{Blank:2018}
V.~D. Blank, V.~D. Churkin, B.~A. Kulnitskiy, I.~A.~Perezhogin, A.~N. Kirichenko,
S.~V. Erohin, P.~B.~Sorokin, and M.~Yu. Popov,
Pressure-Induced Transformation of Graphite and Diamond to Onions,
Crystals {\bf 8}, 68 (2018). 

\bibitem{Shenderova:2002}
O.~A. Shenderova, V.~V. Zhirnov, and D.~W. Brenner,
Carbon Nanostructures,
Critical Reviews in Solid State and Materials Sciences {\bf 27}, 227 (2002). 

\bibitem{Xu:2008}
B.-S. Xu, Prospects and research progress in nano onion-like fullerenes,
New Carbon Materials {\bf 23}, 289 (2008). 

\bibitem{Butenko:2011}
Y.~V. Butenko, L. \v{S}iller, and M.~R.~C. Hunt, Carbon onions,
in: {\it Handbook of nanophysics. Clusters and fullerenes}, Ed. K.~D.~Sattler
(CRC Press: Boca Raton, 2011) 34. 

\bibitem{Chang:2011}
C. Chang, B. Patzer, and D. S\"ulzle, Onion-Like Inorganic Fullerenes,
in: {\it Handbook of nanophysics. Clusters and fullerenes}, Ed. K.~D. Sattler
(CRC Press: Boca Raton, 2011) 51. 

\bibitem{He:2011}
C. He and N. Zhao, Production of Carbon Onions,
in: {\it Handbook of nanophysics. Clusters and fullerenes}, Ed. K.~D. Sattler
(CRC Press: Boca Raton, 2011) 24. 

\bibitem{Bartelmess:2014}
J. Bartelmess and S. Giordani,
Carbon nano-onions (multi-layer fullerenes): chemistry and applications,
Beilstein J. Nanotechnol. {\bf 5}, 1980 (2014). 

\bibitem{Georgakilas:2015}
V. Georgakilas, J.~A. Perman, J. Tucek, and R. Zboril,
Broad Family of Carbon Nanoallotropes: Classification, Chemistry, and
Applications of Fullerenes, Carbon Dots, Nanotubes, Graphene, Nanodiamonds,
and Combined Superstructures,
Chem. Rev. {\bf 115}, 4744 (2015). 

\bibitem{Mykhailiv:2017}
O. Mykhailiv, H. Zubyk, and M.~E. Plonska-Brzezinska,
Carbon nano-onions: Unique carbon nanostructures with fascinating
properties and their potential applications,
Inorganica Chimica Acta {\bf 468}, 49 (2017). 

\bibitem{Butenko:2017}
Y.~V. Butenko, L. \v{S}iller, and M.~R.~C. Hunt, Carbon onions,
in: {\it Nanomaterials Handbook}, Ed. Yu. Gogotsi
(CRC Press: Boca Raton, 2017) 392. 

\bibitem{Muller:1960}
E.~W.~M\"uller, Field ionization and field ion microscopy,
in: {\it ``Advances in Electronics and Electron Physics''},
Ed. L.~Marton (Academic Press: New York, 1960) vol. 13, p. 83. 
  
\bibitem{Forbes:2003}
R.~G.~Forbes, Field electron and ion emission from charged surfaces:
a strategic historical review of theoretical concepts,
Ultramicroscopy {\bf 95}, 1 (2003). 

\bibitem{Fowler:1986}
P.~W. Fowler and J.~Woolrich, $\pi$-Systems in three dimensions,
Chem. Phys. Lett. {\bf 127}, 78 (1986). 

\bibitem{Bochvar:1973}
D.~A. Bochvar and E.~G. Gal'pern, 
On hypothetic systems: carbon-dodehaedron $s$-icosahedron and
carbon-$s$-icosahedron,
Doklady Akad. Nauk USSR {\bf 209}, 610 (1973). 

\bibitem{Saito:1992}
S. Saito, A. Oshiyama, and Y. Miyamoto, 
Electronic structures of fullerenes and fullerides,
in: {\it ``Computational Approaches in Condensed-Matter Physics''},
edited by S. Miyashita, M. Imada, and H. Takayama (Springer: Berlin, 1992) 22.

\bibitem{Verkhovtsev:2013}
A.~V. Verkhovtsev, A.~V. Korol, and A.~V. Solov'yov,
Quantum and classical features of the photoionization spectrum of $C_{60}$,
Phys. Rev. A {\bf 88}, 043201 (2013). 

\bibitem{Jaskolski:1996}
W. Jask\'olski, Confined many-electron systems,
Phys. Rept. {\bf 27}, 1 (1996). 

\bibitem{Lohr:1992}
L.~L. Lohr and M. Blinder, Electron photodetachment from a Dirac bubble
potential. A model for the fullerene negative ion $C_{60}^−$,
Chem. Phys. Lett. {\bf 198}, 100 (1992). 

\bibitem{Puska:1993}
M.~J. Puska and R.~M. Nieminen, Photoabsorption of atoms inside $C_{60}$,
Phys. Rev. A {\bf 47}, 1181 (1993). 

\bibitem{Amusia:1998}
M.~Ya. Amusia, A.~S. Baltenkov, and B.~G. Krakov,
Photodetachment of negative $C_{60}^-$ ions, 
Phys. Lett. A {\bf 243}, 99 (1998). 

\bibitem{Connerade:1999a}
J.~P. Connerade, V.~K. Dolmatov, P.~A. Lakshmi, and S.~T. Manson,
Electron structure of endohedrally confined atoms: atomic
hydrogen in an attractive shell,
J. Phys. B: At. Mol. Opt. Phys. {\bf 32}, L239 (1999). 

\bibitem{Connerade:1999b}
J.~P. Connerade, V.~K. Dolmatov, and S.~T. Manson,
A unique situation for an endohedral metallofullerene,
J. Phys. B: At. Mol. Opt. Phys. {\bf 32}, L395 (1999). 

\bibitem{Schrange:2016}
G. Schrange-Kashenock, $4d\rightarrow 4f$ resonance in photoabsorption of
cerium ion $Ce^{3+}$ and endohedral cerium in fullerene complex $Ce$@$C_{82}^+$,
J. Phys. B: At. Mol. Opt. Phys. {\bf 49}, 185002 (2016). 

\bibitem{Felfli:2018}
Z. Felfli and A.~Z. Msezane,
Simple method for determining binding energies of fullerene negative ions,
Eur. Phys. J. B {\bf 72}, 78 (2018). 

\end{thebibliography}
\end{document}